\documentclass{siamltex}
\usepackage{amssymb}
\usepackage{amsmath}
\usepackage{epsfig}
\usepackage{graphicx}

\title{Computational methods and results for structured multiscale models of tumor invasion}
\author{Bruce P. Ayati\thanks{Department of Mathematics, Southern
Methodist University, Dallas, TX 75205 ({\tt ayati@smu.edu})} \and
Glenn F. Webb\thanks{Department of Mathematics, Vanderbilt
University, Nashville, TN 37240 ({\tt glenn.f.webb@vanderbilt.edu}),
supported by PHS - NIH Grant \#1P50CA113007-01} \and Alexander R. A.
Anderson\thanks{Division of Mathematics, University of Dundee, Dundee
DD1 4HN, Scotland ({\tt anderson@maths.dundee.ac.uk}), supported by
PHS - NIH Grant \#1P50CA113007-01}}

\date{}

\begin{document}

\def\trnum{2005-01}
\def\trauths{Bruce P. Ayati, Glenn F. Webb \\ and Alexander R.A. Anderson}
\def\trtitle{Computational Methods and Results for Structured Multiscale Models of Tumor Invasion}
%
%
%
%

\voffset -.25in

\thispagestyle{empty}
\fontsize{12}{12}
\setcounter{page}{0}

\begin{center}
  \centerline{
  \psfig{figure=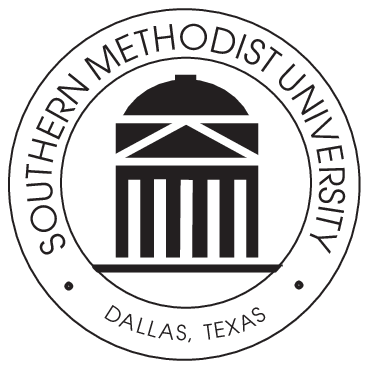,height=1.5in,width=1.5in}
  }

  \vspace{1.25in}

  \begin{minipage}[t]{4in}
    \begin{center}
  
      {\bf \trtitle

      \vspace{24pt}

      \trauths

      \vspace{12pt}

      SMU Math Report \trnum 
      }
    \end{center}
  \end{minipage}

  \vspace{3in}

  {\Large\bf D}{\large\bf EPARTMENT OF} 
  {\Large\bf M}{\large\bf ATHEMATICS} 

  \vspace{3pt}

  {\Large\bf S}{\large\bf OUTHERN} 
  {\Large\bf M}{\large\bf ETHODIST} 
  {\Large\bf U}{\large\bf NIVERSITY} 
\end{center}

\voffset 0in

\normalsize
\newpage

\baselineskip=.33truein
\bigskip

\maketitle
\begin{abstract}
We present multiscale models of cancer tumor invasion with components
at the molecular, cellular, and tissue levels.  We provide biological
justifications for the model components, present computational
results from the model, and discuss the scientific-computing
methodology used to solve the model equations.  The models and
methodology presented in this paper form the basis for developing and
treating increasingly complex, mechanistic models of tumor invasion
that will be more predictive and less phenomenological.  Because many
of the features of the cancer models, such as taxis, aging and
growth, are seen in other biological systems, the models and methods
discussed here also provide a template for handling a broader range
of biological problems.
\end{abstract}



\section{Introduction}

In this paper we present multiscale models of cancer tumor invasion,
and the scientific-computing methodology for solving the model
equations.  The specific model treated here has components at the
molecular level (incorporated via diffusion and taxis processes), the
cellular level (incorporated via a cell age variable),
and the tissue level (incorporated via spatial variables). The tumor
consists of populations of
proliferating and quiescent cells. Proliferating cells are capable of
growing, dividing, entering quiescence, and becoming necrotic.  We consider one mutation class of
proliferating and quiescent cells.  The different physical scales
cause the model to have widely different time scales.  The fully
continuous model treated in detail in this paper depends on variables
representing time, age, and two spatial dimensions.  We present this
system as a simplification of a more general system that depends on
time, age, size, and three spatial dimensions, and has an arbitrary
number of mutation classes for proliferating and quiescent cells,
with increasingly aggressive invasion characteristics.  Mathematical
modeling of all phases of cancer tumor development, angiogenesis, and
metastasis is a very broad and active area of mathematical biology
\cite{AdamBellomo,AndersonChaplain,AraujoMcElwain,cancerSurvey,HornWebb,cancerModelling}.

This paper focuses on the invasion of nearby tissue by a vascular
tumor, under the assumption that the surrounding tissue is the source of the vasculature.  Our fully continuous models have components that are based on
hybrid discrete-continuous (HDC) models
\cite{Sandy2003b,Sandy2003c,SandyHybridCancer,Sandy2000,Sandy2003a}
which use a discrete lattice to represent
space.    We use a physiological variable, age, to model aging in the
proliferating and quiescent tumor cell populations
\cite{DysonWebb,GyllenbergWebb}.  The models in this paper belong to
the class of so-called structured population models in which
individuals in a population are tracked by properties such as age,
size, maturity, and other quantifiable variables. Diffusion and
haptotaxis terms account for the spatial dynamics of the system in
the models under study.  Age, size and/or space structure has also been used in models of tumor cords \cite{BertuzziD'OnofrioFasanoGandolfi,BertuzziFasanoGandolfiMarangi,BertuzziGandolfi,DysonVBWebb}.

Computational and software considerations often limit scientists from
incorporating physiological structure directly into a model.  We
discuss the combination of effective computational methodologies for
integration over the time, age and space variables; we use a
moving-grid Galerkin method for the age variable, an adaptive
step-doubling method for the time variable, and an alternating
direction implicit (ADI) scheme for the space variables.

This paper is organized into three main sections.  The first develops
the models and presents their biological justifications.  The second
section presents computed solutions to the model equations and
discusses their significance.  The third section discusses the
computational methodology.  We close with a section on conclusions
and further research.

\section{Model Equations}
We extend the hybrid discrete continuous (HDC)
tumor invasion model discussed in \cite{SandyHybridCancer} to
fully deterministic models. In particular, as with \cite{SandyHybridCancer}, we focus on four key variables implicated in the invasion process: tumor cells, surrounding tissue (extracellular matrix), matrix-degradative enzymes and oxygen.  Tumor cell motion in the HDC model is driven by a mixture of both biased and unbiased migration where the biased migration is assumed to be from haptotaxis (in response to gradients in the surrounding tissue) and the unbiased migration is just random motility, we shall assume the same here. We assume as in \cite{SandyHybridCancer} that tumor cells produce matrix degrading enzymes which in turn degrade the surrounding tissue creating gradients for the cells to respond to haptotactically. Oxygen production is assumed to be proportional to the tissue density and be consumed by the tumor (see \cite{SandyHybridCancer} and references therein for a more detailed explanation of the HDC model derivation).

One of the important features of the model proposed  in \cite{SandyHybridCancer} was the implementation of tumor heterogeneity i.e. the tumor is made up of many different sub-populations with different phenotypes. These phenotypes allow us to model sub-populations with different invasive capacities. We use the same idea here by considering multiple populations of tumor cells with potentially different parameter values.

Since the models we present here are continuous in all variables, then individual processes of the tumor cells (such as division) are also considered to be continuous. These are modeled according to cell age in the simplified model used in the computations, and cell age and size in
the more general system.   As with the HDC model, these models are
based on the populations of proliferating and quiescent tumor
cells, the density of surrounding tissue macromolecules, the concentration of matrix degradative enzyme, and the concentration of oxygen.

The general class of partial differential equations  for
diffusion and age structure considered in
this paper has a long  history.  Among the
first classic works are Skellam (1951) \cite{skellam} (who
considered the effects of diffusion on populations), and
Sharpe and Lotka (1911) \cite{SharpeLotka} and McKendrick (1926) (who
considered
population models with linear age structure) \cite{McKendrick,Webb}.
More recently, Gurtin and MacCamy \cite{GnM1} considered models with
nonlinear age structure.  Rotenberg \cite{roten} and  Gurtin
\cite{gurtin} posed
models dependent on both age and space.  Gurtin and MacCamy
\cite{GnM2}
differentiated between two kinds of diffusion in these models:
diffusion due to random dispersal, and diffusion toward an area of
less crowding.  Existence
and uniqueness results can be found for various forms of these models
in  Busenberg and Iannelli
\cite{BnI}, di Blasio \cite{diblasio}, di Blasio and Lamberti
\cite{DnL}, Langlais \cite{langlais1}, MacCamy
\cite{maccamy}, and Webb \cite{Webb80}. Further analysis has been
done by several authors
\cite{huang,KnL,langlais2,marcati}.

\subsection{The Age-, Space- and Size-structured model of Tumor
Invasion}
\begin{figure}[ht]
\begin{center}
\includegraphics[width=4.0in]{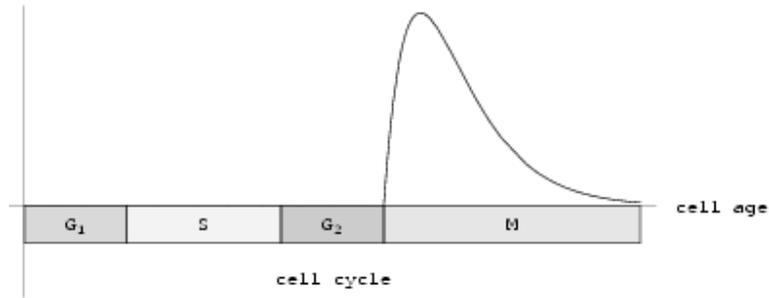}
\caption{Schematic of the phases $G_{1}$ (first gap), $S$ (synthesis), $G_{2}$ (second gap), and $M$ (mitosis) of the cell cycle
correlated to cell age. The graph over the mitotic phase corresponds
to the distribution of cell ages at division (the response of the function $\theta$ to age.)}
\label{cell.cycle}
\end{center}
\end{figure}

The tumor is contained in a region of tissue $\Omega$. The tumor is
composed of proliferating cells (cells that are transiting
the cell cycle to mitosis) and quiescent cells (cells that
are arrested in the cycle, but are capable of resuming
progress). We assume that proliferating cells are motile in
space, but quiescent cells are not, and that both proliferating
and quiescent cells consume oxygen, with quiescent cells at
a lower rate (as in \cite{SandyHybridCancer}). Cells, both proliferating and quiescent, are distinguished
by their position $x \in \Omega$, their age $a$ between $0$
(newly divided) and $a_{M}$ (maximum possible
age), their size $s$ between $s_{m}$ (minimum possible size) and
$s_{M}$ (maximum possible size), and their state in the mutation sequence. Cell age, for both proliferating cells and quiescent cells,
is the time since the cell was newly divided. For
proliferating cells, cell age correlates to phase of the cell cycle
(first gap $G_{1}$, synthesis $S$, second gap $G_{2}$, and mitosis).
An illustration of a distribution of division ages is given in
Figure \ref{cell.cycle}. Cells are also distinguished by cell size,
which  can be interpreted as mass, diameter, volume, or
other measurable property. The inclusion of cell age and cell size allows
description of the growth of the tumor mass to be understood at the
level of individual cells, as they double their size and divide
to two new daughter cells.  For example, the inclusion of age and size in the diffusivities represents a means by which growing and dividing cells increase total tumor size.

In the HDC model in \cite{SandyHybridCancer}, the behavior
of individual cells is tracked cell by cell on a spatial lattice. This
discrete formulation relates detailed information about fundamental
processes at the cellular level, such as cell-cell adhesion,
entry to and from quiescence, division, apoptosis, and phenotype
mutation, to behavior of the tumor mass. In the
continuous age-size structured model of this paper,
behavior at the population level is also related to behavior at the
individual cell level, with cell age and size dependent densities
providing the connection to these processes.  The use of continuous
densities constitutes a local averaging of individual traits.

The dependent variables of the model are:

\begin{itemize}
\item $p_{i}(x,a,s,t)$ = density of proliferating tumor cells of type
$i$ in the tumor at
position $x$, age $a$, and size $s$ at time $t$, where $i=0$
corresponds to a mutated type p53 gene, and $i=1,2,\dots,n$
corresponds to a linear sequence of
mutated phenotypes of increasing aggressiveness. The number of
mutations can be very large, with successive phenotypes possessing
greater proliferative characteristics and capacity for spatial movement.

\item $q_{i}(x,a,s,t)$ = density of quiescent tumor cells of type $i$
in the tumor at
position $x$, age $a$, size $s$, and mutation phenotype $i=0,2,\dots,n$
at time $t$.

\item $f(x,t)$ = surrounding tissue macromolecule (MM) density at
position $x$ at time $t$. It is assumed that these macromolecules are
distributed heterogeneously in $\Omega$, but immobile in $\Omega$.

\item $m(x,t)$ = matrix degradative enzyme (MDE) concentration at
position $x$ at time $t$. MDE is produced by the tumor cells and
diffuses in $\Omega$.

\item $c(x,t)$ = oxygen concentration at position $x$ at time $t$.
Oxygen is produced by the extracellular MM, diffuses in
$\Omega$, and is consumed by the tumor cells.

\item $P(x,t) = \sum_{i=0}^{n} \, \int_{0}^{a_{M}}
\int_{s_{m}}^{s_{M}}
\, p_{i}(x,a,s,t) \,ds \, da =$ the total population density in $x$
of proliferating cells
of all types at time $t$.

\item $Q(x,t) = \sum _{i=0}^{n} \, \int _{0}^{a_{M}}
\int_{s_{m}}^{s_{M}} \, q_{i}(x,a,s,t) \, ds \, da =$ the total
population density in $x$ of quiescent cells of all types at time $t$.

\item $N(x,t) = P(x,t) + Q(x,t)=$ total tumor population density in
  $x$ of all cell types at time $t$.
\end{itemize}

The equations governing the proliferating-cell densities of the tumor
are
\begin{subequations}
\begin{align}\frac{\partial}{\partial t} p_{i}(x,a,s,t) = &
- \, \underbrace{\frac {\partial}{\partial a} p_{i}(x,a,s,t)}
_{\mbox{\tiny cell aging}} \, - \,
\underbrace{\frac {\partial}{\partial s} (\kappa_{i} (a,s,c)
p_{i}(x,a,s,t))}_{\mbox{\tiny cell growth}} \label{p_i} \\
&+
\underbrace{
\nabla \cdot (D_{p_{i}}(x,a,s,N) \nabla
p_{i}(x,a,s,t))}_{\mbox{\tiny diffusion}} - \underbrace{\chi_{i}
\nabla \cdot (p_{i}(x,a,s,t)
\nabla f(x,t))}_{\mbox{\tiny haptotaxis}} \nonumber \\
& -
\underbrace{\rho_{i}(x,a,s,c,N) p_{i}(x,a,s,t)}_{\mbox{\tiny cell death
    from insufficient oxygen}} - \underbrace{\theta_{i}(x,a,s,c,N)
p_{i}(x,a,s,t)}_{\mbox{\tiny division with sufficient oxygen}}
\nonumber \\
& - \underbrace{\sigma_{i}(x,a,s,c,N) p_{i}(x,a,s,t)}_
{\mbox{\tiny exit to quiescence}}
+ \underbrace{\tau_{i}(x,a,s,c,N) q_{i}(x,a,s,t)}_{\mbox{\tiny entry
from
quiescence}}, \nonumber
\end{align}

\noindent with age-boundary conditions
\begin{align}
\underbrace{p_{i}(x,0,s,t)}_{\mbox{\tiny newborn type $i$ cells}} = &
\, 4 (1 \, -\psi_{i} \,) \, \underbrace{\int_{0}^{a_M}
\theta_{i}(x,a,2s,c,N(x,t)) p_{i}(x,a,2s,t) \, da}_
{\mbox{\tiny type $i$ cell division}} \label{p_birth} \\
&+ \, 4 \, \psi_{i-1}  \, \underbrace{\int_{0}^{a_M}
\theta_{i-1}(x,a,2s,c,N(x,t))
 p_{i-1}(x,a,2s,t) \, da}_
{\mbox{\tiny type $i-1$ cell division}}, \nonumber
\end{align}

\noindent where $\psi_i$ is the fraction of type $i$ cells with type
$i+1$ mutation.  For cells that have undergone only one
primary cancer forming mutation (such as a p53 mutation), we
set  $i=0$ and $\psi_{-1}=0$.  The coefficient of 4, rather than the
more intuitive splitting value of 2, results from the assumption of
even cell division; uneven cell division would require a mitosis
kernel and integration over the size variable, $s$, in equation
(\ref{p_birth}) \cite{TuckerNZimmerman,Webb89}.

The equations governing the quiescent-cell densities are
\begin{align}
\frac {\partial}{\partial t} q_{i}(x,a,s,t) =  & -
\underbrace{\frac {\partial}{\partial a} q_{i}(x,a,s,t)}
_{\mbox{\tiny cell aging}} -
\underbrace{\nu_{i}(x,a,s,c,N(x,t)) q_{i}(x,a,s,t)}_{\mbox{\tiny cell death
    from insufficient oxygen}} \label{q_i} \\
&+
\underbrace{\sigma_{i}(x,a,s,c,N(x,t)) p_{i}(x,a,s,t)}_{\mbox{\tiny
entry from proliferation}} -
\underbrace{\tau_{i}(x,a,s,c,N(x,t)) q_{i}(x,a,s,t)}_{\mbox{\tiny exit to
    proliferation}}. \nonumber
\end{align}
The quiescent-cell populations lack a boundary condition in age since
they are "born" when proliferating cells of the same mutation class
become quiescent.

The equations governing tissue macromolecule, matrix degradative
enzyme, and oxygen densities are precisely those used in
\cite{SandyHybridCancer}:
\begin{align}
\frac {\partial}{\partial t} f(x,t) \, = \,& - \,
\underbrace{\delta m(x,t) f(x,t)}_{\mbox{\tiny degradation}},
\label{f} \\
\frac {\partial}{\partial t} m(x,t) \, = \, &
\underbrace{D_{m} \nabla ^{2} m(x,t)}_{\mbox{\tiny diffusion}}
\, + \, \underbrace{\mu P(x,t)}_{\mbox{\tiny production}}
\, - \, \underbrace{\lambda m(x,t)}_{\mbox{\tiny decay}}, \label{m} \\
\frac{\partial}{\partial t} c(x,t) \, = \, &
\underbrace{D_{c} \nabla ^{2} c(x,t)}_{\mbox{\tiny diffusion}}
\, + \, \underbrace{\beta f(x,t)}_{\mbox{\tiny production}} \, - \,
\underbrace{\gamma P(x,t)-\eta Q(x,t)}_{\mbox{\tiny uptake}} \,
\, - \, \underbrace{\alpha c(x,t)}_{\mbox{\tiny decay}}. \label{c}
\end{align}
\end{subequations}
\noindent Equations (\ref{p_i})-(\ref{c}) are combined with initial
conditions and
no-flux boundary conditions on the boundary $\partial \Omega$ of
$\Omega$.

Equation (\ref{p_i}) balances the way cells age, grow, and move in time.
The first term on the right-side of equation (\ref{p_i}) accounts for
the aging of cells, which is one-to-one with advancing time. In the second
term in equation (\ref{p_i}), $\kappa_{i}(a,s,c)$ is the rate at which
proliferating
cells increase size, i.e., $\int_{s_{1}}^{s_{2}}
\frac{1}{\kappa_{i}(a,s,c)}ds$ is the time required for a cell of type
$i$ to grow
from size $s_{1}$ to size $s_{2}$. The diffusion term in equation
(\ref{p_i}) accounts for cell movement due to random
motility, interphase drag, the interaction between cells,
volume displacement due to cell division, and cell-cell adhesion
\cite{AraujoMcElwain}.  The diffusion
coefficient $D_{p_{i}}(x,a,s,N(x,t))$ can be allowed to
depend on the independent and dependent
variables to incorporate mechanistic features of these processes.
For example, cells in higher mutation phenotype classes may have
smaller cell-cell adhesion properties, and thus have a larger coefficient.
Dividing cells of larger size may exert
greater force of volume displacement, and thus have a larger
coefficient.
In equation (\ref{p_i}), the haptotaxis term represents directed
movement of cells toward concentrations of MM, which is the source of
oxygen necessary for tumor cell growth, and is degraded by tumor cell
produced MDE.
The coefficient $\rho_{i}(x,a,s,c,N(x,t))$ of
proliferating cell loss in equation (\ref{p_i}) is dependent
on the density of cells in competition for the supply of oxygen.
In equation (\ref{p_birth}), $\theta_{i}(x,a,s,c,N(x,t))$ is the
rate at which cells of type $i$, age $a$, and size $s$ divide at $x$
per unit time, where it is assumed that a mother cell divides into two
daughter cells of equal size (unequal division can also be modeled
\cite{Webb89}). The division rate $\theta_{i}(x,a,s,c,N(x,t))$
depends on the age of cells, the supply of oxygen,
as well as on the density of cells, with reduced capacity
for division as the oxygen supply decreases and the density increases.
The negative sign in front of $\theta_{i}(x,a,s,c,N(x,t))$ reflects
the loss of cells due to the division process. The mother cell of age
$a$ and size $s$ is replaced by two daughter cells, each having age
$0$ and half the size of the mother cell, as described in the
boundary condition (\ref{p_birth}).
The coefficients $\sigma_{i}(x,a,s,c,N(x,t))$ and $\tau_{i}(x,a,s,c,N(x,t))$
of transition to and from quiescence in equation (\ref{p_i})
depend on the supply of oxygen and the density of tumor cells. Lower
oxygen and higher density results in increased entry to quiescence
and higher oxygen and lower density results in increased recruitment
from quiescence.
The equation (\ref{q_i}) governing the quiescent cells is interpreted
similarly, where it is assumed that quiescent cells are not motile.
In this model, we represent the properties of individual cell behavior
as rates of transition dependent on cell spatial position, age, and
size. The inclusion of cell age and size
structure allows incorporation of cell level processes without
tracking of each cell history, cell by cell (as is done in \cite{SandyHybridCancer}). The hybrid and continuum
modeling approaches have complementarity in development,
analysis, and computability, in which advantages of each can be
exploited.

\subsection{A Simplified Two-Dimensional Model with No Size Structure}
The following model is a version of the model above with no size
structure, two spatial dimensions (denoted by $(x,y)\in \Omega$), and
one compartment each of
proliferating- and quiescent-cell types.  The equations governing the
two classes of cell densities of the tumor are
\begin{subequations}
\begin{align}\frac{\partial}{\partial t} p(x,y,a,t) =  &
-  \underbrace{\frac {\partial}{\partial a} p(x,y,a,t)}
_{\mbox{\tiny cell aging}} \\ & +
\underbrace{D_{p}\nabla ^{2} p(x,y,a,t)}_{\mbox{\tiny diffusion}}
- \underbrace{\chi \nabla \cdot \big( p(x,y,a,t) \nabla f(x,y,t)
\big)}_{\mbox{\tiny haptotaxis}} \label{p} \\
& -
\underbrace{\rho(x,y,a,c) p(x,y,a,t)}_{\mbox{\tiny cell death
    from insufficient oxygen}} - \underbrace{\theta(x,y,a,c)
p(x,y,a,t)}_{\mbox{\tiny division with sufficient oxygen}} \nonumber
\\
& -  \underbrace{\sigma(x,y,a,c,N(x,t)) p(x,a,s,t)}_
{\mbox{\tiny exit to quiescence}}
+  \underbrace{\tau(x,y,a,c) q(x,y,a,t)}_{\mbox{\tiny entry from
quiescence}}, \nonumber \\
\frac {\partial}{\partial t} q(x,y,a,t) =  & -
\underbrace{\frac {\partial}{\partial a} q(x,y,a,t)}
_{\mbox{\tiny cell aging}} -
\underbrace{\nu(x,y,a,c) q(x,y,a,t)}_{\mbox{\tiny cell death
    from insufficient oxygen}} \label{q} \\
&+
\underbrace{\sigma(x,y,a,c,N(x,t)) p(x,y,a,t)}_{\mbox{\tiny entry
from proliferation}} -
\underbrace{\tau(x,y,a,c) q(x,y,a,t)}_{\mbox{\tiny exit to
    proliferation}}, \nonumber
\end{align}
\noindent with age-boundary conditions
\begin{align}
\underbrace{p(x,y,0,t)}_{\mbox{\tiny newborn cells}} \, = \, 2
\underbrace{\int_{0}^{a_0} \theta(x,y,a,c) p(x,y,a,t) \ da}_
{\mbox{\tiny division rate}}. \label{ps_birth}
\end{align}
\end{subequations}

The equations governing tissue macromolecule ($f$), matrix degradative
enzyme ($m$), and oxygen ($c$) densities remain as defined in
equations
(\ref{f})-(\ref{c}).  All equations are combined with initial
conditions and
zero flux boundary conditions on an $(x,y)$-rectangle $\Omega$.

\section{Computations of Cancer Tumor Invasion}

We can demonstrate some aspects of the behavior of the reduced system
defined by equations (\ref{p})-(\ref{ps_birth}) and equations
(\ref{f})-(\ref{c}) through computations using
parameters and functional forms chosen for illustrative purpose
rather than biological foundation.  Take the spatial domain
$\Omega=[-5,5]\times[-5,5]$ and take
\begin{subequations}
\begin{equation} \label{paramStart}
D_p = 0.0005, \,
\chi = 0.01, \,
D_m = 0.01, \,
D_c = 0.05,
\end{equation}
\begin{equation}
\rho(x,y,c) = 0.1 \, \max\{1.0 - c, 0\}, \,
\nu(x,y,c) = 2.0 \, \max\{1.0 - c, 0\},
\end{equation}
\begin{equation}
\delta(x,y) = 50.0, \,
\mu(x,y) = 1.0,   \,
\lambda(x,y) = 0.0, \,
\beta(x,y) = 0.5,
\end{equation}
\begin{equation}
\gamma(x,y) = 0.57, \,
\eta(x,y) = 0.0, \,
\alpha(x,y) = 0.025,
\end{equation}
\begin{equation}
\sigma(x,y,c) = 10.0 \, \max\{1.0 - c, 0\}, \,
\tau(x,y,c) = 2.0 \, c.
\end{equation}

The distribution of division ages is assumed to have the
form of an offset integrand of the Gamma function (see Figure \ref{cell.cycle}),
\begin{equation}
\theta(x,y,a,c) = \left\{ \begin{array}{rl} 10.0 \, c \exp(-10(a-1))
\ (2a-1)^5, &a > 0.5,  \\ 0, &a<0.5, \end{array} \right.
\end{equation}
where 0.5 is the minimum age at which a cell can divide.  The initial
conditions are
\begin{eqnarray}
p(x,y,a,0) &=& 5.0 \, G(\sqrt{(x-5)^{2}+(y-5)^{2}},0,0.5),\\
q(x,y,a,0) &=& 0.5 \, p(x,y,0),\\
f(x,y,0) &=& 0.2 \, \cos(0.4 \, x^{2}) \, \sin(0.2\,y^{2})+0.2,\\
m(x,y,0) &=& 2.5 \, G(\sqrt{(x-5)^{2}+(y-5)^{2}},0,0.5), \\
c(x,y,0) &=& 10.0 \, f(x,y,0), \label{paramEnd}
\end{eqnarray}
\end{subequations}
\noindent where
$$G(z,z_\mu,z_\sigma) = \frac{\exp(-\frac{(z-z_\mu)^{2}}{2 \,
z_\sigma^{2}})}{\sqrt{2 \, \pi} \, z_\sigma}.$$

\begin{figure}[t]
\begin{center}
\epsfig{file=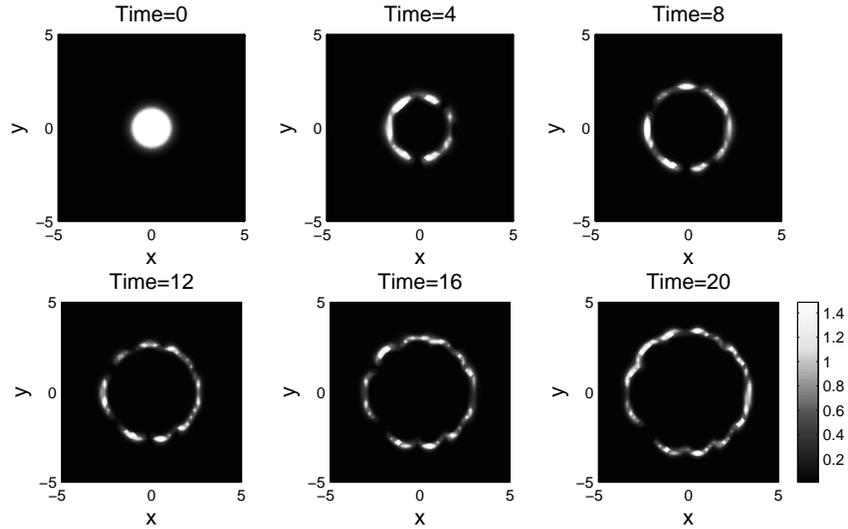,width=5.0in}
\caption{Proliferating-cell density ($P$) for the system defined by equations
(\ref{p})-(\ref{ps_birth}) and equations (\ref{f})-(\ref{c}) .  The
parameters used in this computation are defined in equations
(\ref{paramStart})-(\ref{paramEnd}).}
\label{outputP}
\end{center}
\end{figure}

\begin{figure}[t]
\begin{center}
\epsfig{file=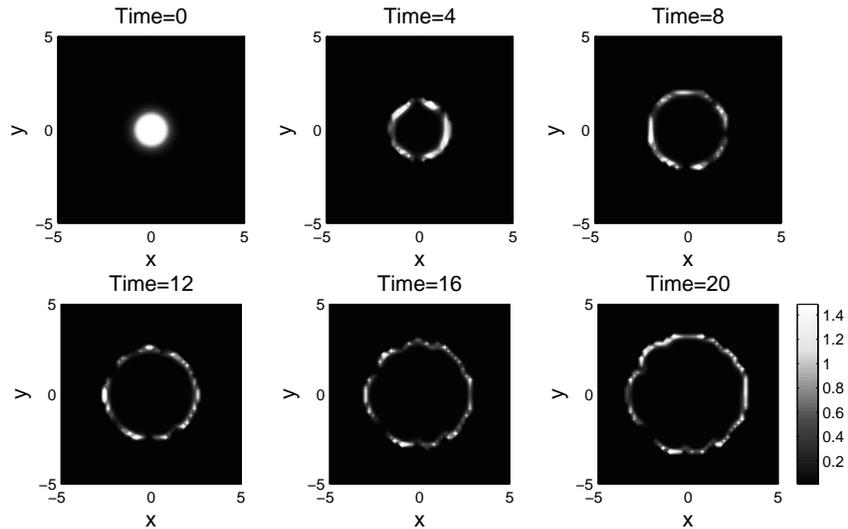,width=5.0in}
\caption{Quiescent-cell density ($Q$) for the system defined by equations
(\ref{p})-(\ref{ps_birth}) and equations (\ref{f})-(\ref{c}) .  The
parameters used in this computation are defined in equations
(\ref{paramStart})-(\ref{paramEnd}).}
\label{outputQ}
\end{center}
\end{figure}

\begin{figure}[t]
\begin{center}
\epsfig{file=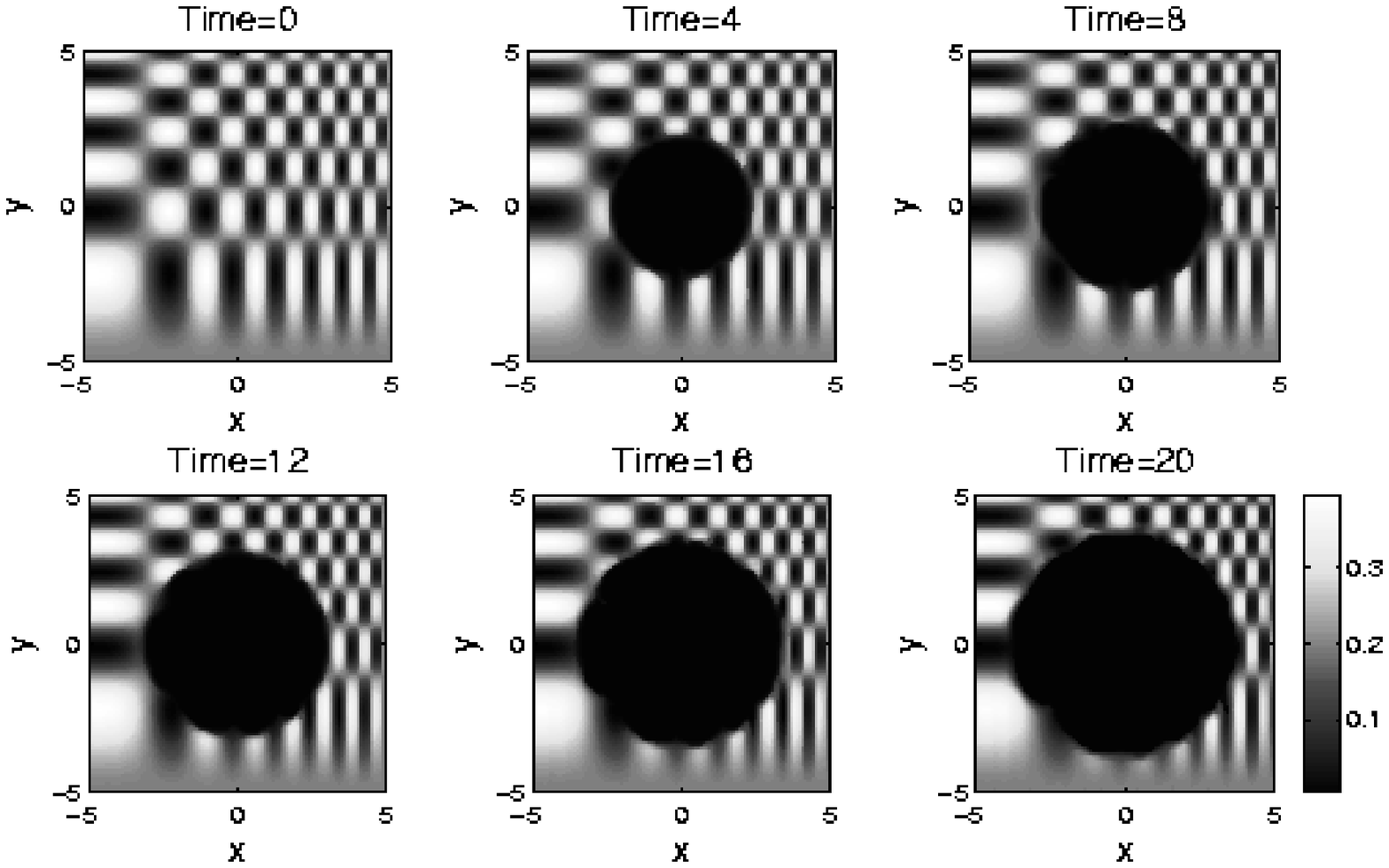,width=5.0in}
\caption{Macromolecule density ($f$) for the system defined by equations
(\ref{p})-(\ref{ps_birth}) and equations (\ref{f})-(\ref{c}) .  The
parameters used in this computation are defined in equations
(\ref{paramStart})-(\ref{paramEnd}).}
\label{outputf}
\end{center}
\end{figure}

Numerical computations of the proliferating-cell density and macromolecule density for the simplified model are illustrated in Figures
\ref{outputP}-\ref{outputf} as snapshots in time\footnote{Animations can be found
online at http://faculty.smu.edu/ayati/cancer.html}.  The
simulation in Figures \ref{outputP}-\ref{outputf} demonstrates the temporal development of
spatial heterogeneity in the tumor mass from a radially symmetric
initial condition of tumor cells and heterogeneous initial condition
of surrounding macromolecules. The macromolecule tissue is displaced
by the tumor tissue as a consequence of haptotactic movement of the
tumor cells, driven by the matrix degradative enzyme they produce.
The interior core of the tumor mass becomes necrotic, because of its
increasing distance from the oxygen supply provided by the
macromolecule matrix source.  Once aspect of this computation is that the tumor edge consists of an outer layer of proliferating cells and an inner layer of quiescent cells.

\section{Computational Methodology}

Computational robustness and efficiency is vital for the methods used
to solve the
high-dimension, multiscale models developed in this paper.  The
primary issue is the age discretization and how to decouple it from
the time discretization without ignoring the fact that age and time
advance together.  This approach foreshadows how one may wish to
handle size structure.  The decoupling of the age and time
discretizations allows for adaptivity in the time variable; we
discuss a particularly effective method for the time integration
called step-doubling.  The third computational consideration is in
how we solve the system in the space variables.  We use an
alternating direction implicit method, which is a novel approach when
incorporated into the step-doubling method for time.

There is a plethora of numerical methods for solving
models with just age or size structure
\cite{AnguloL-M99,AnguloL-M04,chiu,FnL-M,InKnP,KnC,L-M,sulsky}.
These methods use uniform age and timesteps which are equal to one
another in the case of age structure, or do the equivalent in the
context of size structure of introducing a new size node at every
time step.  This approach does not work well
for problems with multiple time scales
because the fastest time scale tends to be in the spatial variables.

To understand the nature of this problem, consider a fixed, uniform
age discretization.  Solving the system along characteristics would
require the age interval width equal to the time step.  This would
result in many more age nodes than are needed to accurately solve the
problem in the age variable because of the small time step.  For size
structure, the analogous situation is to introduce many more size
nodes at the birth boundary than are needed.  An additional concern
with size structure is that characteristic curves in the size-time
plane can converge, resulting in unnecessarily narrow size
intervals.  Regridding was used in \cite{sulsky} and
\cite{AnguloL-M04} to adjust for the effects of narrowing gaps
between characteristics, but they do not address the issue of small
size nodes at the birth boundary.  For example, the method proposed
in \cite{AnguloL-M04} has an advantage of simplicity -- the idea is
to merge the narrowest size interval with one of its neighbors after
each time step -- but is not a satisfactory solution because small
size intervals can arise continuously at the birth boundary while
elsewhere size intervals continue to narrow due to the nature of the
characteristic curves.  Moreover, regridding comes at a computational
cost.  A natural solution to this problem lies in using a finite
element space with a moving reference frame in age or size, which is
the approach we use in this paper.

Previous numerical methods designed explicitly for models with
dependence on age, time and space were developed outside the context
of an application and required uniform age and time discretizations
with the age step chosen to equal the time step \cite{kim,KnP,LnT}.
In contrast, the methods used to obtain the computational results
presented in this paper
\cite{age-pwconst-paper,age-general-paper} were motivated by models
of {\em Proteus mirabilis} swarm-colony development where the need to
decouple age and time discretizations was clear from the problem
\cite{proteus-tech,EnS,rauprich}.  In the process of applying these
methods to the system defined in \cite{EnS}, it became clear that the
numerical methods and software used previously were not merely
inefficient, but also gave qualitatively incorrect answers (although
these methods did decouple the age and time discretizations, they did
so by not moving the age discretization along characteristics; see
the appendix in \cite{proteus-tech} for a discussion.)  This is a
critical pitfall to avoid and highlights the importance of using
methods with known convergence properties for a particular system.

We use Galerkin finite element methods that use a moving grid to
allow for independent, nonuniform age and time discretizations and
whose development has focused on robustness as well as computational
efficiency.  The important property of these methods is that the age
step need not equal the time
step. Instead, the positions of the age nodes are adjusted by the time
step. The methods preserve the important fact that age and time
advance together.  The methods in \cite{kim,KnP,LnT} also discretized
along
characteristics, but they did so simultaneously in age and time and
thus imposed the often
crippling constraint that the time and age steps be both constant and
equal.   The difficulty with this approach is twofold.  First, the
use of constant age and time
steps prevents adaptivity of the discretization in age or,
especially, time.   Second, and more
importantly, the coupling of the age and time meshes can cause great
losses of efficiency since only rarely will the dynamics in time be on
the same scale as the dynamics in age.  This is particularly the case
when space is involved since sharp moving fronts can require small
time steps,
whereas the behavior in the age variable can remain relatively
smooth. The age discretization presented in \cite{kim,KnP,LnT} can be
viewed as special cases of the methods
presented in \cite{age-pwconst-paper,age-general-paper} by setting
the time and age
meshes to be constant and equal and using a backward Euler
discretization in time and a piecewise constant finite element space
in age.

Step-doubling \cite{step-doubling-paper,gear,shampine85} is a
conceptually simple, yet quite effective method for the adaptive time
integration of differential equations.  Over a time step, we compute
one solution over the entire time step, and then a second solution
over two successive half steps.  These two different solutions give
us two things.  First, we can compare solutions to determine the
accuracy of our approximation for the purposes of adaptivity of the
time step.  Second, we can combine solutions to get a likely
second-order accurate approximation, even when each step in the
step-doubling process is first-order accurate.

To solve the model equations in the spatial variables, we use an ADI
method (also called operator splitting) where we first solve the
equations in just the $x$ derivatives and zero-order terms, and then
in just the $y$ derivatives
\cite{DouglasDupontADI,KarlsenSplitting,McLachlanQuispel,StrangSplitting,ThomasVol1}.

This approach reduces our two-dimensional problem in space to a set
of more easily solved one-dimensional spatial problems; we need to
solve a series of block tridiagonal linear systems instead of a more
computationally expensive wide-banded linear system.  Because ADI
methods are time ordered, the ADI method needs to be embedded into
the step-doubling algorithm.

\begin{figure}[t]
\begin{center}
\includegraphics[width=2.5in]{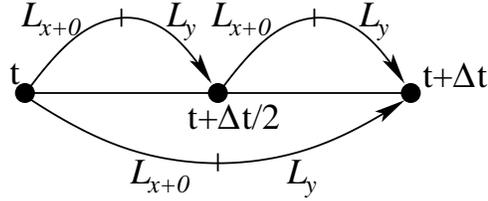}
\caption{Schematic of the combination of the step-doubling and ADI
methods to advance the solution of a time and space dependent system
from time $t$ to time $t+\Delta t$.  The operators $L_{x+0}$ and
$L_y$ represent the $x$ derivatives plus zero-order terms, and the
$y$ derivatives, respectively.}
\label{step-doubling-ADI}
\end{center}
\end{figure}

The combination of these methods results in the following breakdown
of the model equations.  First, the moving-grid Galerkin methods in
age reduce the age-, time-, and space-dependent equations to systems
of differential equations that depend on time and two spatial
variables.  We then solve each of these equations by a combination of
step-doubling and ADI methods; we take a step in the $x$ direction
and zero-order terms, followed by a step in the $y$-direction, within
each substep of step-doubling.  This integrated stepping is
illustrated in Figure \ref{step-doubling-ADI}.

The software used to generate the computed solutions in this paper
has a similar structure to BuGS \cite{BuGS}.  The age methods
presented in \cite{age-pwconst-paper,age-general-paper} use
discontinuous piecewise polynomials as basis function for the age
space, which results in a distinct system of parabolic partial
differential equations for each age interval if we keep time and
space continuous.  This in turn results in a distinct linear system
for each age interval when we fully discretize the equations.  In the
tumor invasion software, we use piecewise constant functions in age
with post-processing to
continuous piecewise linear functions.  As mentioned above, the tumor
invasion software works by updating the age discretization at the
beginning of a time step and then applying the step-doubling method
to the subsystems corresponding to each age interval splitting the
spatial operator into two separate operators over each dimension.

As in BuGS, the tumor invasion software requires the user to define
the spatial discretization of the equations by writing a residual
function based on first-order backward differences in time.  The
software then uses the implementation of the step-doubling method
described in \cite{step-doubling-paper} to get a second-order
accurate in time implicit finite difference scheme.  The software also
features step control for the convergence of Newton's method and
automatic approximation of the Jacobi matrix.

\section{Conclusions and Further Research}

In this paper we presented physiologically and spatially structured
continuous deterministic models of cancer tumor invasion.  We
presented a general model whose equations depend on variables
representing size, age, space and time.  We then treated a simplified
model without size structure and with only two spatial dimensions.
The simplified model contained one mutation class of proliferating
and quiescent cells.  The aim of this approach is to move tumor
invasion modeling away from phenomenological models toward more
mechanistic,  biologically informed, and reliably predictive models.
These more complex models required a
more sophisticated computational methodology to investigate
numerically the computationally intensive model equations.

The most immediate extension of this work is to determine the models
parameters and functional forms from biological data and
experiments.  The current methodology and software is sufficient to
handle multiple mutation classes of proliferating and quiescent
cells, but a deeper understanding of the biology is needed to benefit
from this extension.  Computational results from more biologically
detailed models are expected, in turn, to contribute to a deeper
understanding of the underlying biology.

The most important mathematical extension of the methodology is to
develop size-time finite elements to handle size-structured
equations.  Rather than being developed for general forms of
transport, extensions
of the existing methods for age structure to size structure will use
the specific nature of physiological
change in tumor cells to allow the incorporation of size structure
into a model at a low cost in terms of computational resources.
Anticipated complications in handling size structure include birth in
a size-structured context with respect to both the numerical methods
and their analyses.  Since the
characteristic curves in the size-time plane are no longer lines with
slope one, as was the case for age structure, some important
questions are: what types of characteristic curves should we consider
and how do
we handle situations where these curves become asymptotically close
within the moving grid framework?  What happens if they meet and
shocks form?

Two immediate concerns must be addressed for the
problem of size dependence in tumor invasion models.  The first issue
is the introduction of new size nodes at the birth boundary, and the
second is the handling of size intervals that contract due to the
convergence of size-time characteristic curves.  We expect the major
complication in the size nodes to occur when growth slows as cells
reach a certain size.  However, because of the nonlinearities in the
problem, it is insufficient to merely assume that a size interval
will strictly decrease length.  Addressing these two concerns will
lay the foundation for methods that handle more complicated
characteristics, including the formation of shocks that can form in
situations where growth has complex dependencies on the physiological
traits of an individual as well as the external environment.

As in the methods for age-structured systems, the moving grid
formulation is expected to account for the growth of individuals,
taking the place of direct differencing of the size variable.  And as
in the case of age structure, the use of a space of discontinuous
piecewise polynomials as the basis functions in size is expected to
allow each size interval to be treated with a separate linear
system.  If the system has dependence on both age and size, we would
have a two dimensional array of independent linear systems at each
time step.

An important benefit of using size-time Galerkin finite elements is
having one mathematical framework define many methods with
higher-order accuracy.  Because of the need to keep computational
costs down in each dimension of the high-dimension systems under
study, without sacrificing robustness, the ability to choose the
order of convergence of the method is quite useful.

A major extension of the software and methodology is to add a third
space dimension through an additional sub-operator in the ADI
method.  This methodology for handling three space dimensions is
expected to be sufficient for generating initial results that aim to
extend our understanding of tumor invasion beyond the 2D-space
models.  Other ADI methods that may work within this framework are
Douglas-Gunn \cite{ThomasVol1} and Strang Splitting
\cite{StrangSplitting}.

Although we have provided a specific mathematical treatment of the
spatial dynamics of tumor invasion, we remark that modeling spatial
dynamics can be more complicated in biological systems than in
physical systems.  A broad examination of
different modeling approaches is required, including the continuous
approach
 in this paper, and how it relates to other approaches, such as the
hybrid discrete-continuous formulation discussed in
\cite{SandyHybridCancer}.  Multiscale models of the type considered
in this paper have different time scales for the dynamics at the
different physical scales.  For example, in the system defined in
equations (\ref{p_i})-(\ref{c}), the cellular scale gives rise to
time scales in the age and size variables, whereas the tumor scale
gives rise to a different time scale in the spatial variables.
Independent of the specific type of spatial representation used,
decoupling time from age or size is critical for effective solution
of the model equations.

Many of the features of the cancer models, such as taxis, aging and
growth, are seen in other biological systems; prior work on {\em
Proteus mirabilis} swarm-colony development is but one example
\cite{proteus-tech}.  Biological systems abound where either spatial
dynamics induce the behavior of interest, or where the spatial
dynamics is the behavior of interest.  In the same manner, the
behavior of interest in a biological system can depend on the
distribution of physiological traits such as age or size, or those
distributions are the topic of interest.  We hope that the
methodology presented in this paper will provide a template for
handling a broader range of biological problems.

\newpage

\bibliographystyle{siam}
\bibliography{cancer,age,na,math,bio,apps}

\end{document}